# ACCRETION INTO BLACK HOLES WITH MAGNETIC FIELDS, AND RELATIVISTIC JETS

Bisnovatyi-Kogan G.S., Klepnev A.S.
Space Research Institute Rus. Acad. Sci., Moscow, Russia
Lovelace R.V.E.
Cornell University, Ithaca, USA

**Abstract**

We discuss the problem of the formation of a large-scale magnetic field in the accretion disks around black holes, taking into account the non-uniform vertical structure of the disk. The high electrical conductivity of the outer layers of the disk prevents the outward diffusion of the magnetic field. This implies a stationary state with a strong magnetic field in the inner parts of the accretion disk close to the black hole, and zero radial velocity at the surface of the disk. Structure of advective accretion disks is investigated, and conditions for formation of optically thin regions in central parts of the accretion disk are found. The problem of jet collimation by magneto-torsion oscillations is considered.

## 1. Introduction

Quasars and AGN contain supermassive black holes, about 10 HMXR contain stellar mass black holes - microquasars. Jets are observed in objects with black holes: collimated ejection from accretion disks.

Early work on disk accretion to a black hole argued that a large-scale magnetic field of, for example, the interstellar medium would be dragged inward and greatly compressed by the accreting plasma [10,11,13]. Subsequently, analytic models of the field advection and diffusion in a turbulent disk suggested, that the large-scale field diffuses outward rapidly [13,15], and prevents a significant amplification of the external poloidal field. The question of the advection/diffusion of a large-scale magnetic field in a turbulent plasma accretion disk was reconsidered in [7], taking into account its non-uniform vertical structure. The high electrical conductivity of the surface layers of the disk, where the turbulence is suppressed by the radiation flux and the high magnetic field, prevents outward diffusion of the magnetic field. This leads to a strong magnetic field in the inner parts of accretion disks.

The standard model for accretion disks [18] is based on several simplifying assumptions. The disk must be geometrically thin and rotate at the Kepler angular velocity. These assumptions make it possible to neglect radial gradients and, to proceed from the differential to algebraic equations. For low accretion rates $\dot{M}$, this assumption is fully appropriate. However, for high accretion rates, the disk structure may differ from the standard model. To solve the more general problem, advection and a radial pressure gradient have been included in the analysis of the disk structure [17]. It was shown in [1], that for large accretion rates there are no local solutions that are continuous over the entire region of existence of the disk and undergo Kepler rotation. A self-consistent solution for an advective accretion disk with a continuous description of the entire region between the optically thin and optically thick regions had been obtained in [3,6].

## 2. The fully turbulent model

There are two limiting accretion disk models which have analytic solutions for a large-scale magnetic field structure. The first was constructed in [11] for a stationary non-rotating accretion disk. A stationary state in this disk (with a constant mass flux onto a black hole) is maintained by the balance between magnetic and gravitational forces, and thermal balance (local) is maintained by Ohmic heating and radiative heat conductivity for an optically thick conditions. The mass flux to the black hole in the accretion disk is determined by the finite electrical conductivity of the disk matter and the diffusion of matter across the large-scale magnetic field. It is widely accepted that the laminar disk is unstable to different hydrodynamic, magneto-hydrodynamic, and plasma instabilities which implies that the disk is turbulent. In X-ray binary systems the assumption about turbulent accretion disk is necessary for construction of a realistic models [18]. The turbulent accretion disks had been constructed for non-rotating models with a large-scale magnetic field. A formula for turbulent magnetic diffusivity was derived in [11], similar to the scaling of the shear $\alpha$-viscosity in turbulent accretion disk in binaries [18], where the viscous stress tensor component $t_{r\phi} = \alpha P$, with $\alpha \leq 1$ a dimensionless constant, and $P$ is the pressure in the disk midplane. Using this representation, the expression for the turbulent electrical conductivity $\sigma_t$ is written as

$$\sigma_t = \frac{c^2}{\tilde{\alpha} 4\pi h \sqrt{P/\rho}}. \tag{1}$$

Here, $\tilde{\alpha} = \alpha_1 \alpha_2$. The characteristic turbulence scale is $\ell = \alpha_1 h$, where $h$ is the half-thickness of the disk, the characteristic turbulent velocity is $v_t = \alpha_2 \sqrt{P/\rho}$. The large-scale magnetic field threading a turbulent Keplerian disk arises from external electrical currents and currents in the accretion disk. The magnetic field may become dynamically important, influencing the accretion disk structure, and leading to powerful jet formation, if it is strongly amplified during the radial inflow of the disk matter. It is possible only when the radial accretion speed of matter in the disk is larger than the outward diffusion speed of the poloidal magnetic field due to the turbulent diffusivity $\eta_t = c^2/(4\pi\sigma_t)$. Estimates in [15] have shown that for a turbulent conductivity (1), the outward diffusion speed is larger than the accretion speed. Thus it appears that there is no large-scale magnetic field amplification during Keplerian disk accretion. Numerical calculations in [15] are reproduced analytically for the standard accretion disk structure [7]. Far from the inner disk boundary the specific angular momentum is $j \gg j_{in}$. The characteristic time $t_{visc}$ of the matter advection due to the shear viscosity is $t_{visc} = \frac{r}{v_r} = \frac{j}{\alpha v_s^2}$. The time of the magnetic field diffusion is $t_{diff} = \frac{r^2}{\eta} \frac{h}{r} \frac{B_z}{B_r}$, $\eta = \frac{c^2}{4\pi\sigma_t} = \tilde{\alpha} h v_s$. In the stationary state, the large-scale magnetic field in the accretion disk is determined by the equality $t_{vis} = t_{diff}$, what determines the ratio

$$\frac{B_r}{B_z} = \frac{\alpha}{\tilde{\alpha}} \frac{v_s}{v_K} = \frac{\alpha}{\tilde{\alpha}} \frac{h}{r} \ll 1. \qquad (2)$$

Here, $v_K = r\Omega_K$ and $j = rv_K$ for a Keplerian disk. In a turbulent disk a matter is penetrating through magnetic field lines, almost without a field amplification: the field induced by the azimuthal disk currents has $B_{zd} \sim B_{rd}$.

## 3. Turbulent disk with radiative outer zones

Near the surface of the disk, in the region of low optical depth, the turbulent motion is suppressed by the radiative flux, similar to the suppression of the convection over the photospheres of stars with outer convective zones. The presence of the outer radiative layer does not affect the estimate of the characteristic time $t_{visc}$ of the matter advection in the accretion disk because it is determined by the main turbulent part of the disk. The time of the field diffusion, on the contrary, is significantly changed, because the electrical current is concentrated in the radiative highly conductive regions, which generate the main part of the magnetic field. Inside the turbulent disk the electrical current is negligibly small so that the magnetic field there is almost fully vertical, with $B_r \ll B_z$. In the outer radiative layer, the field diffusion is very small, so that matter advection is leading to strong magnetic field amplification. We suppose, that in the stationary state the magnetic forces could support the optically thin regions against gravity. When the magnetic force balances the gravitational force in the outer optically thin part of the disk of surface density $\Sigma_{ph}$, the following relation takes place [11]

$$\frac{GM\Sigma_{ph}}{r^2} \approx \frac{B_z I_\phi}{2c} \approx \frac{B_z^2}{4\pi}, \qquad (3)$$

The surface density over the photosphere corresponds to a layer with effective optical depth close to $2/3$ (e.g. [5]). We estimate the lower limit of the magnetic field strength, taking $\kappa_{es}$ (instead of the effective opacity $\kappa_{eff} = \sqrt{\kappa_{es}\kappa_a}$). Writing $\kappa_{es}\Sigma_{ph} = 2/3$, we obtain $\Sigma_{ph} = 5/3$ (g/cm$^2$) for the opacity of the Thomson scattering, $\kappa_{es} = 0.4$ cm$^2$/g. The absorption opacity $\kappa_a$ is much less than $\kappa_{es}$ in the inner regions of a luminous accretion disk so we estimate the lower bound on the large-scale magnetic field in a Keplerian accretion disk as [7]

$$B_z = \sqrt{\frac{5\pi}{3}} \frac{c^2}{\sqrt{GM}} \frac{1}{x\sqrt{m}} \approx 10^8 \, \text{G} \frac{1}{x\sqrt{m}}, \quad x = \frac{r}{r_g}, \quad m = \frac{M}{M_\odot}. \qquad (4)$$

The maximum magnetic field is reached when the outward magnetic force balances the gravitational force on the surface with a mass density $\Sigma_{ph}$. In equilibrium, $B_z \sim \sqrt{\Sigma_{ph}}$. We find that $B_z$ in a Keplerian accretion disk is about $20$ times less than its maximum possible value [11], for $x = 10, \alpha = 0.1,$ and $\dot{m} = 10$.

## 4. Self-consistent numerical model

Self-consistent models of the rotating accretion disks with a large-scale magnetic field requires solution the equations of magneto-hydrodynamics. The solution with a small field will not be stationary, and a transition to the strong field solution will take place. Therefore the strong field solution is the only stable stationary solution for a rotating accretion disk. The vertical structure of the disk with a large scale poloidal magnetic field was calculated in [14], taking into account the turbulent viscosity and diffusivity, and the fact that the turbulence vanishes at the surface of the disk. Coefficients of the turbulent viscosity $\nu$, and magnetic diffusivity $\eta$ are connected by the magnetic Prandtl number $P \sim 1$, $\nu = P\eta = \alpha \frac{c_{s0}^2}{\Omega_K} g(z)$, where $\alpha$ is a constant, determining the turbulent viscosity [18]; $\beta = c_{s0}^2/v_{A0}^2$, where $v_{A0} = B_0/(4\pi\rho_0)^{1/2}$ is the midplane Alfvén velocity. The function $g(z)$ accounts for the absence of turbulence in the surface layer of the disk [7]. In the body of the disk $g = 1$, whereas near the surface of the disk $g$ tends over a short distance to a very small value, effectively zero. The smooth function with a similar behavior is taken [15] in the form $g(\zeta) = \left(1 - \frac{\zeta^2}{\zeta_s^2}\right)^\delta$, with $\delta \ll 1$. In the stationary state the boundary condition on the disk surface is $u_r = 0$, and only one free parameter - magnetic Prandtl number $P$ remains in the problem. In a stationary disk vertical magnetic field has a unique value. The example of the radial velocity distribution for $P = 1$ is shown in Fig.1 from [8].

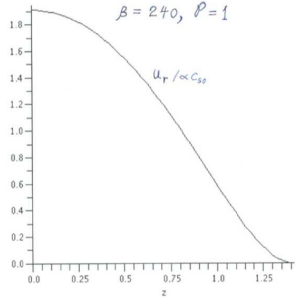

Fig. 1. Distribution of the radial velocity over the thickness in the stationary accretion disk with a large scale poloidal magnetic field

## 5. Basic equations for accretion disk structure

We use equations describing a thin, steady-state accretion disk, averaged over its thickness [3,6]. These equations include advection and can be used for any value of the vertical optical thickness of the disk. We use a pseudo-newtonian approximation for the structure of the disk near the black hole, where the effects of the general theory of relativity are taken into account using the Paczycski-Wiita potential [16]

$$\Phi(r) = -\frac{GM}{r - 2r_g}. \tag{5}$$

Here $M$ is the mass of the black hole, $2r_g = 2GM/c^2$ is the gravitational radius. The self-gravitation of the disk is neglected, the viscosity tensor $t_{r\phi} = -\alpha P$. The conservation of mass is expressed in the form $\dot{M} = 4\pi r h \rho v$, where $\dot{M}$ is the accretion rate, $\dot{M} > 0$, and $h$ is the half thickness of the disk. The equilibrium in the vertical direction $\frac{dP}{dz} = -\rho z \Omega_K^2$ is replaced by the algebraic relation in the form $h = \frac{c_s}{\Omega_K}$, where $c_s = \sqrt{P/\rho}$ is the isothermal sound speed. The equations of motion in the radial and azimuthal directions are, respectively, written as

$$v\frac{dv}{dr} = -\frac{1}{\rho}\frac{dP}{dr} + (\Omega^2 - \Omega_K^2)r, \quad \frac{\dot{M}}{4\pi}\frac{d\ell}{dr} + \frac{d}{dr}(r^2 h t_{r\phi}) = 0, \tag{6}$$

where $\Omega_K$ is the Kepler angular velocity, given by $\Omega_K^2 = GM/r(r - 2r_g)^2$; $\ell = \Omega r^2$ is the specific angular momentum. Other components of the viscosity tensor are assumed negligibly small. The vertically averaged equation for the energy balance is $Q_{adv} = Q^+ - Q^-$, where

$$Q_{adv} = -\frac{\dot{M}}{4\pi r}\left[\frac{dE}{dr} + P\frac{d}{dr}\left(\frac{1}{\rho}\right)\right], \quad Q^+ = -\frac{\dot{M}}{4\pi}r\Omega\frac{d\Omega}{dr}\left(1 - \frac{l_{in}}{l}\right), \tag{7}$$

$$Q^- = \frac{2aT^4 c}{3(\tau_\alpha + \tau_0)h}\left[1 + \frac{4}{3(\tau_0 + \tau_\alpha)} + \frac{2}{3\tau_*^2}\right]^{-1}, \tag{8}$$

are the energy fluxes (erg/cm$^2$/s) associated with advection, viscous dissipation, and radiation from the surface,

respectively, $\tau_0$ is the Thomson optical depth, $\tau_0 = 0.4\rho h$ for the hydrogen composition. We have introduced the optical thickness for absorption, $\tau_\alpha \approx 5.2 \cdot 10^{21} \dfrac{\rho^2 T^{1/2} h}{acT^4}$, and the effective optical thickness $\tau_* = \left[(\tau_0 + \tau_\alpha)\tau_\alpha\right]^{1/2}$. The equation of state is for a mixture of a matter and radiation $P_{tot} = P_{gas} + P_{rad}$. The gas pressure is given by formula, $P_{gas} = \rho RT$, $R$ is the gas constant, and the radiation pressure is given by

$$P_{rad} = \frac{aT^4}{3}\left[1 + \frac{4}{3(\tau_0 + \tau_\alpha)}\right]\left[1 + \frac{4}{3(\tau_0 + \tau_\alpha)} + \frac{2}{3\tau_*^2}\right]^{-1}. \qquad (9)$$

The specific energy of the mixture of the matter and radiation is determined as $\rho E = \dfrac{3}{2} P_{gas} + 3 P_{rad}$. Expressions for $Q^-$ and $P_{rad}$, valid for any optical thickness, have been obtained in [1].

### 6. Method of solution and numerical results

The system of differential and algebraic equations can be reduced to two ordinary differential equations,

$$\frac{x}{v}\frac{dv}{dx} = \frac{N}{D}, \qquad (10)$$

$$\frac{x}{v}\frac{dc_s}{dx} = 1 - (\frac{v^2}{c_s^2} - 1)\frac{N}{D} + \frac{x^2}{c_s^2}(\Omega^2 - \frac{1}{x(x-2)^2}) + \frac{3x-2}{2(x-2)}. \qquad (11)$$

Here the numerator $N$ and denominator $D$ are algebraic expressions depending on $x, v, c_s$, and $l_{in}$, the equations are written in dimensionless form with $x = r/r_g$, $r_g = GM/c^2$. The velocities $v$ and $c_s$ have been scaled by the speed of light $c$, and the specific angular momentum $l_{in}$ by the value $c/r_g$. This system of differential equations has two singular points, defined by the conditions $D = 0$, $N = 0$. The inner singularity is situated near the last stable orbit with $r = 6r_g$. The outer singularity, lying at distances much greater than $r_g$, is an artifact arising from our use of the artificial parametrization $t_{r\phi} = -\alpha P$ of the viscosity tensor. The system of ordinary differential equations was solved by a finite difference method discussed in [2]. The method is based on reducing the system of differential equations to a system of nonlinear algebraic equations which are solved by an iterative Newton-Raphson scheme, with an expansion of the solution near the inner singularity and using of $l_{in}$ as an independent variable in the iterative scheme [2]. The solution is almost independent of the outer boundary condition. The numerical solutions have been obtained for the structure of an accretion disk over a wide range of the parameters $\dot{m}$ ($\dot{m} = \dfrac{\dot{M}c^2}{L_{EDD}}$) and $\alpha$. For low accretion rates, $\dot{m} < 0.1$, the solution for the advection model has $\delta_* \gg 1$, $v \ll c_s$, and an angular velocity is close to the Kepler velocity everywhere, except a very thin layer near the inner boundary of the disk. As the accretion rate increases, the situation changes significantly. The changes show up primarily in the inner region of the disk. Fig. 2 shows the radial dependences of the temperature of the accretion disk for the accretion rate $\dot{m} = 50$, and different values of the viscosity parameter $\alpha = 0.01, 0.1, 0.4$. Clearly, for large $\dot{m}$ and $\alpha$ the inner part of the disk becomes optically thin. Because of this, a sharp increase in the temperature of the accretion disk is observed in this region.

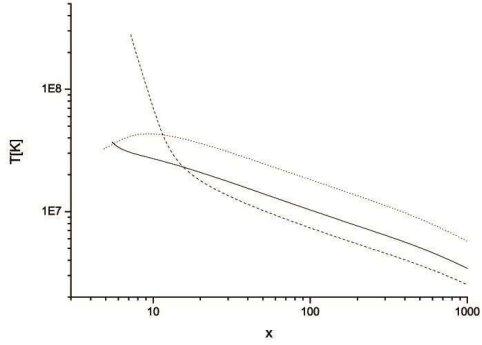

Fig. 2. The radial dependence of the temperature of the accretion disk for an accretion rate $\dot{m} = 50$, and viscosity parameters $\alpha = 0.01$ (dotted curve), $\alpha = 0.1$ (smooth curve), and $\alpha = 0.4$ (dashed curve).

Two distinct regions can be seen in the plot of the radial dependence of the temperature of the accretion disk. This is especially noticeable for a viscosity parameter $\alpha = 0.4$, where one can see the inner optically thin region with a dominant non-equilibrium radiation pressure $P_{rad}$, and an outer region which is optically thick with dominant equilibrium radiation pressure. Things are different when the viscosity parameter is small. Only a small (considerably smaller than for $\alpha = 0.4$) inner region becomes optically thin for accretion rates of $\dot{m} \approx 30 - 70$. Meantime, in the case of $\alpha = 0.01$, there are no optically thin regions at all.

### 6. Jet collimation by magneto-torsional oscillations.

Following [4], we consider the stabilization of a jet by a pure magneto-hydrodynamic mechanism associated with torsional oscillations. We suggest that the matter in the jet is rotating, and different parts of the jet rotate in different directions, see Fig.3. Such a distribution of the rotational velocity produces an azimuthal magnetic field, which prevents a disruption of the jet. The jet is representing a periodical, or quasi-periodical structure along the axis, and its radius oscillates with time all along the axis. The space and time periods of oscillations depend on the conditions at jet formation: the length-scale, the amplitude of the rotational velocity, and the strength of the magnetic field. The time period of oscillations can be obtained during the construction of the dynamical model, and the model should also show at what input parameters a long jet stabilized by torsional oscillations could exist.

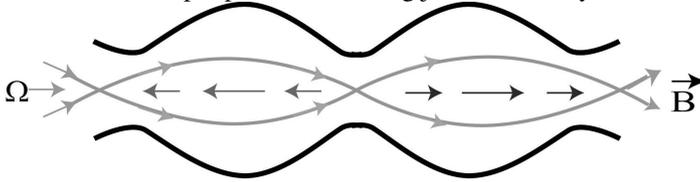

Fig.3. Qualitative picture of jet confinement by magneto-torsional oscillations.

Let us consider a long cylinder with a magnetic field directed along its axis. This cylinder will expand without limit under the action of pressure and magnetic forces. It is possible, however, that a limiting value of the radius of the cylinder could be reached in a dynamic state, in which the whole cylinder undergoes magneto-torsional oscillations. Such oscillations produce a toroidal field, which prevents radial expansion. There is therefore competition between the induced toroidal field, compressing the cylinder in the radial direction, and gas pressure, together with the field along the cylinder axis (poloidal), tending to increase its radius. During magneto-torsional oscillations there are phases when either the compression or expansion force prevails, and, depending on the input parameters, there are three possible kinds of behavior of such a cylinder that has a negligible self-gravity.
(1) The oscillation amplitude is low, so the cylinder suffers unlimited expansion (no confinement).
(2) The oscillation amplitude is high, so the pinch action of the toroidal field destroys the cylinder and leads to the formation of separated blobs.
(3) The oscillation amplitude is moderate, so the cylinder, in absence of any damping, survives for an unlimited time, and its parameters (radius, density, magnetic field etc.) change periodically, or quasi-periodically, in time.
After considerable simplifications, which details may be found in [4], the equation, describing the magneto-torsional oscillations of a long cylinder, takes the following form:

$$\frac{d^2 y}{d\tau^2} = \frac{1 - D\sin^2 \tau}{y}. \qquad (12)$$

This equation describes approximately the time dependence of the outer radius of the cylinder $R(t)$ in the symmetry plane, where the rotational velocity remains zero. The dimensionless variables and the parameter $D$ in (12) are defined as $\tau = \omega t$, $y = \dfrac{R}{R_0}$, with $R_0 = \dfrac{\sqrt{K}}{\omega}$, $D = \dfrac{1}{2\pi K C_m}\left(\dfrac{C_b \Omega_0}{z_0 \omega}\right)^2$. The frequency of oscillations $\omega$ may be represented as $\omega = \alpha_n k V_A = \alpha_n \dfrac{B_{z,0}}{z_0}\sqrt{\dfrac{\pi}{\rho_0}}$, where $k$ is the wave number, $k = 2\pi/z_0$, and $V_A$ is the Alfven velocity, $V_A = B_{z,0}/\sqrt{4\pi\rho_0}$; $\alpha_n < 1$ is a coefficient determining the frequency of nonlinear Alfven oscillations, which are similar to the magneto-torsional oscillations under investigation. The example of the dynamically stabilized cylinder is given in Fig.4, from [4], $y$ and $z$ are non-dimensional radius, and radial velocity, respectively. Transition to stochastic regime in these oscillations was investigated in [9],

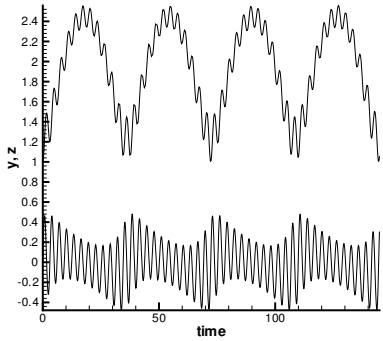

**Fig.4.** Time dependence of non-dimensional radius $y$ (upper curve), and non-dimensional velocity $z$ (lower curve), for $D = 2.1$, $y(0)=1$.

**Acknowledgments**

The work of GSBK, ASK was partially supported by the Russian Foundation for Basic Research grant 11-02-00602, the RAN Program 'Origin, formation and evolution of objects of Universe' and President Grant for Support of Leading Scientific Schools NSh-3458.2010.2. R.V.E.L was supported in part by NASA grants NNX08AH25G and NNX10AF63G and by NSF grant AST-0807129.